\documentclass[floatfix,twocolumn,showpacs,amsmath,amssymb,superscriptaddress]{revtex4-1}
\usepackage{times}
\usepackage{latexsym}
\usepackage{graphicx}
\usepackage{verbatim,times,bbm}
\usepackage{color}
\usepackage{appendix}
\usepackage{amsmath}

\usepackage{color}

\newcommand{\rom}[5]{\expandafter\@slowromancap\romannumeral #5@}

\def\b{\beta}
\def\g{\gamma}

\def\r{\rho}
\def\s{\sigma}

\def\be{\begin{equation}}
\def\ee{\end{equation}}
\def\ba{\begin{eqnarray}}
\def\ea{\end{eqnarray}}
\def\la{\langle}
\def\ra{\rangle}
\def\nn{\nonumber}

\def\rp{\right)}
\def\lp{\left(}
\def\v{\vert}

\begin{document}
\title{Noise Effects on Entanglement Distribution by Separable State}
\author{Najmeh Tabe Bordbar}
\email{email:najmeh@physics.sharif.edu }
\author{Laleh Memarzadeh}
\email{Corresponding author. email: memarzadeh@sharif.ir}
\affiliation{Department of Physics, Sharif
University of Technology, Teheran, Iran}

\begin{abstract}
We investigate noise effects on the performance of entanglement distribution by separable state. We consider a realistic situation in which the mediating particle between two distant nodes of the network goes through a noisy channel. For a large class of noise models we show that the average value of distributed entanglement between two parties is equal to entanglement between particular bipartite partitions of target qubits and exchange qubit in intermediate steps of the protocol. This result is valid for distributing two qubit/qudit and three qubit entangled states. In explicit examples of the noise family, we show that there exists a critical value of noise parameter beyond which distribution of distillable entanglement is not possible. Furthermore, we determine how this critical value increases in terms of Hilbert space dimension, when distributing $d$-dimensional Bell states. 
\end{abstract}

\pacs{03.65.Yz , 03.67.Bg}

\date{\today}

\maketitle
\section{Introduction}\label{intro}
The main obstacle in realization of quantum information tasks is sensitivity of quantum systems to noise \cite{Zurek}. Within  computation or communication processes, any quantum system experiences noise which is due to the interaction between the system and its surrounding environment. As a result of these unavoidable interactions, quantum properties of the systems are disturbed or even destroyed. Error correcting codes and quantum feedback control schemes \cite{Error} are known as powerful tools for reliable communication or storage of information. In addition to that, for realization of any quantum information task and designing successful experiments, it is essential to analyse the noise effects on given protocols. Amongst the most important protocols are those designed for construction of quantum networks \cite{QNetworks}. As quantum network developments are based on reliable entanglement distribution between network nodes, it is of utmost importance to inspect the possible effects of noise on any proposals of entanglement distribution.
\\
\\
\begin{figure}
 \centering
\includegraphics[scale=0.32]{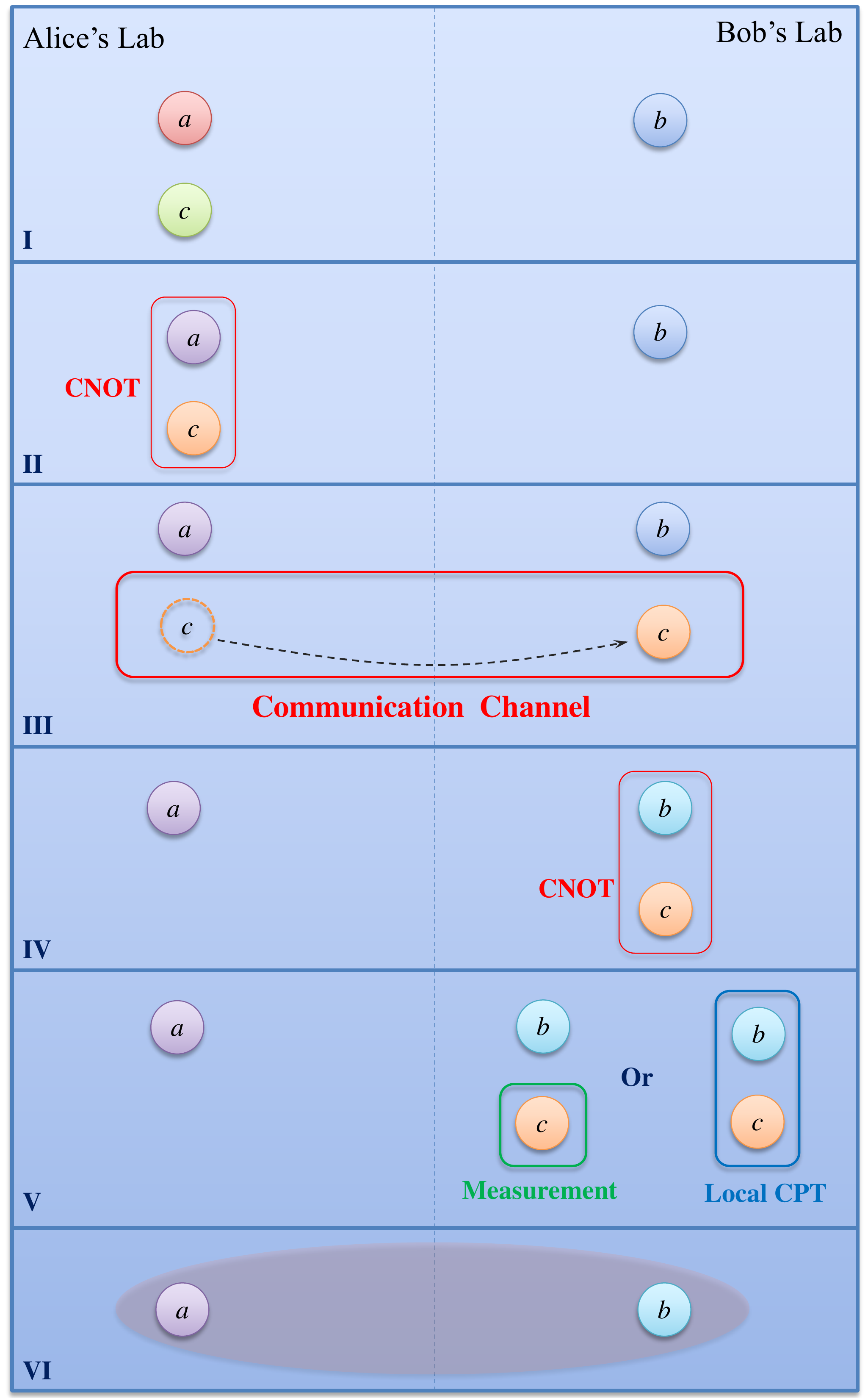}
\caption{Steps of entanglement distribution by separable states. Alice and Bob want to make their qubits $a$ and $b$ entangled while they are far from each other. In step II Alice interacts her qubit with ancilla $c$. In step III Alice sends the ancilla qubit to Bob. Bob interacts his qubit with the ancilla qubit received from Alice (step IV). In step V Bob either measures ancilla qubit $c$ or performs a local channel on qubits $b$ and $c$ to make qubits $a$ and $b$ entangled.}
\label{Presentation1}
\end{figure}
While entanglement can be generated between particles in an isolated laboratory, entanglement distribution between distant nodes of a network is challenging because by local operations and classical communication (LOCC) entanglement can not be generated \cite{Bennet}. 
There are different approaches for entanglement distribution using different sorts of resources and methods \cite{direct1, single1, single2}. One interesting example of such approaches is introduced in  \cite{edss} which is known as entanglement distribution by separable states (EDSS). In this approach (depicted in figure \ref{Presentation1}) no initial entanglement in the system is required. Alice and Bob who are in distant labs have qubits $a$ and $b$ respectively. The aim is to make qubits $a$ and $b$ entangled. One ancillary qubit labelled by $c$ is required as mediating particle, through which qubits $a$ and $b$ interact with each other. The initial state of three qubits $a$, $b$ and $c$ is separable and during all the steps of the protocol shown in figure \ref{Presentation1}, qubit $c$ remains in separable state with qubits $a$ and $b$. However, at the end of the protocol, either maximally entangled state is shared between $a$ and $b$ with non-zero probability (If Bob measures qubit $c$ at step V of figure \ref{Presentation1}) or a deterministic approach can be followed to generate entanglement with value less than one ( If Bob performs a completely positive trace preserving map (CPT map) on qubits $b$ and $c$ which are in his lab at step V  of figure \ref{Presentation1}). This method of entanglement distribution has recently been successfully tested experimentally {\cite{expedss1}} and is generalized to continuous-variable entanglement distribution {\cite{mista1}}.  Systematic method of the protocol for distributing $d$-dimensional Bell states and GHZ states is introduced in {\cite{sedss}}. It is also shown that difference between entanglement in partition $b|ac$ at stage IV of figure (\ref{Presentation1}) and entanglement in partition $a|bc$ at stage II of figure (\ref{Presentation1}) is bounded by quantum discord between qubit $c$ and qubits $ab$ at state IV of the figure (\ref{Presentation1})
\cite{BrussPiani}. This result is generalized to the case when qubit $c$ goes through a noisy channel \cite{Lewenstein}. Furthermore when initial resource of entangled pair is available in one lab, sharing that entanglement by sending one part through the noisy channel is studied in \cite{Pal}. \\
\\
Here we are interested in actual amount of distillable entanglement that can be distributed between distant labs in EDSS protocol (no initial entanglement is required) when the exchanged particles go through noisy channels. For a large and important class of quantum channels, we show that, average value of distillable entanglement distributed between qubits $a$ and $b$ is equal to the distillable entanglement remained in bipartite partition $a|bc$ and $b|ac$ after noise affects the protocol. In other words, although the exchange qubit is in separable state with the target qubits during the whole protocol, distributing entanglement between qubits $a$ and $b$ is possible if entanglement in partitions $a|bc$ and $b|ac$ does not vanish when $c$ goes through the noisy channel. We extend our studies to the case of entanglement distribution between three distant labs by analysing GHZ state distribution when exchange qubits are transferred through noisy channels. We discuss the role of distillable entanglement between different partitions in intermediate steps, for successful entanglement distribution between distant labs. We also study the performance of EDSS protocol for sharing $d$- dimensional two partite entangled states in presence of noise (see appendix \ref{dBell-SEDSS & Noise}).
\\
\\
The structure of this paper is as follows: In sec \ref{EDSS & Separable ancilla} we set our notation and briefly review EDSS protocol. Section \ref{QuantumChannels} is devoted to parametrization of qubit quantum noisy channels. In section \ref{Noise2-qubit}  we study the effect of noise on the performance of the EDSS protocol for distributing two-qubit entangled states both in probabilistic and deterministic approaches. Effect of noise on distributing GHZ states between three distant qubits is studied in section \ref{General Noise on GHZ}. Conclusions will be drawn in section \ref{Summary and Conclusion}.
\section{Entanglement Distribution by Separable States} \label{EDSS & Separable ancilla}
In this section we review the main steps of EDSS protocol which was first introduced in {\cite{edss}}. We set our notation as follows:
\begin{itemize}
\item The order of parties on the right hand side of each equation, follows the order of labels in the left hand side of that equation. For example $\rho_{x,y}=\Omega\otimes\sigma$ is equivalent with $\rho_{x,y}=\Omega_x\otimes\sigma_y$, or $|\psi\rangle_{xyz}=|\phi\rangle|0\rangle$ is equivalent with $|\psi\rangle_{xyz}=|\phi\rangle_{xy}|0\rangle_z$
\item $n$ partite GHZ state in $d$ dimensional Hilbert space is denoted by $|\mathrm{GHZ}_n^{(d)}\rangle$:
\begin{equation*}
|\mathrm{GHZ}_n^{(d)}\rangle=\frac{1}{\sqrt{d}}\sum_{i=0}^{d-1}|i\rangle^{\otimes n}.
\end{equation*}
\item Projective operators are denoted by $\Pi_{i_1,i_2,\cdots i_n}=|i_1,i_2,\cdots ,i_n\ra\la i_1,i_2,\cdots ,i_n|$.
\item CNOT operator with control qudit $x$ and target qudit $y$, is represented by $C_{xy}$. Its action on two qudits is given by: $C_{x,y}|i,j\rangle_{x,y}=|i,j+i\rangle$ . Similar notation is applied for its inverse, $C^{(-1)}_{x,y}$which is defined by $C^{(-1)}_{x,y}|i,j\rangle_{x,y}=|i,j-i\rangle$. All the sums are in mode $ d $.
\end{itemize} 
By using this notation, EDSS protocol for distributing entanglement between two qubits $a$ and $b$ which are respectively in Alice's and Bob's distant labs, is described in the following:\\
\\
Initial state of the protocol $\r_{abc}^{(0)}$ (step I in figure (\ref{Presentation1})), is a separable state which is described by:
\be \label{eq1}
\r_{abc}^{(0)}=\frac{1}{6}\sum_{k=0}^{3}\v\psi_{k},\psi_{-k},0\ra\la\psi_{k},\psi_{-k},0\v+\frac{1}{6}\sum_{i=0}^{1}\v i,i,1\ra\la i,i,1\v,
\ee
where $\v\psi_{k}\ra=\frac{1}{\sqrt{2}}\lp\v 0\ra+e^{\frac{ik\pi}{2}}\v 1\ra\rp $. Alice performs CNOT gate on qubit $a$ and ancilla qubit $c$ which is initially in her lab (step II in figure (\ref{Presentation1})). This action results the following state:
\be \label{eq:firstcnot}
\r_{abc}^{(1)}=C_{ac}(\r_{abc}^{(0)})=\frac{1}{3}\v \mathrm{GHZ_3^{(2)}}\ra\la\mathrm{GHZ}_3^{(2)}\v+\sum_{i,j,k=0}^{1}\b_{ijk}\Pi_{ijk},
\ee
where $\beta_{ijk}=\frac{1}{6}$ if $j\neq k$, otherwise it is zero. 
After applying CNOT, Alice sends qubit $c$ to Bob through an ideal communication channel (step III in figure (\ref{Presentation1})). Bob applies CNOT gate to qubits $b$ and $c$ (step IV in figure (\ref{Presentation1})) which gives the following three-qubit state:
\be
\r_{abc}^{(2)}=C_{bc}(\r_{abc}^{(1)})=\frac{1}{3}\v\psi^{+}\ra\la\psi^{+}\v\otimes\v 0\ra\la 0\v+\frac{1}{6}\mathrm{I}_2\otimes \mathrm{I}_2\otimes\v 1\ra\la 1\v,
\ee
where $\mathrm{I_2} $ is two by two identity matrix and $|\psi^+\rangle=\frac{1}{\sqrt{2}}(|00\ra+|11\ra)$ is a maximally entangled state. The final step for entanglement distribution between qubits $a$ and $b$, is either performing a measurement on qubit $c$, or performing a quantum channel on qubits $b$ and $c$ by Bob (step V in figure (\ref{Presentation1})). \\
\\
If Bob measures qubit $c$ in computational basis, $|\psi^+\rangle $  is shared between $a$ and $b$ with probability $\frac{1}{3} $ and with probability $\frac{2}{3}$ the protocol is unsuccessful as separable state is shared between qubits $a$ and $b$. Hence on average entanglement distributed between Alice and Bob is equal to $\frac{1}{3}$.\\
\\
To avoid probabilistic effects, instead of measurement, Bob can perform local quantum channel on qubits $b$ and $c$ given by
\begin{equation}\label{LCPTMAP}
\Phi_{bc}(\r)=\sum_{j=1}^3A_{bc}^{(j)}\r A_{bc}^{(j)\dagger},
\end{equation}
with Kraus operators:
\begin{equation*} 
A_{bc}^{(1)}=\mathrm{I}_2\otimes\v 0\ra\la 0\v ,\quad A_{bc}^{(2)}=|01\ra\la 01|  ,\quad A_{bc}^{(3)}=|01\ra\la 11|.
\end{equation*}
Final state of qubits $a$ and $b$ after the action of the channel $\Phi_{bc}$ and tracing over qubit $c$ is given by:
\be\label{eqlcpt}
\r_{ab}=\mathrm{tr}_{c}(\Phi_{bc}(\r_{abc}^{(2)}))=
\frac{1}{3}\v\psi^{+}\ra\la\psi^{+}\v+\frac{1}{3}\mathrm{I}\otimes |0\ra\la 0|.
\ee
Entanglement shared between qubits $a$ and $b$ quantifying by concurrence \cite{concurrence} (see appendix \ref{concurrenceDef} for concurrence definition) is equal to $\frac{1}{3}$. It is worth noticing that if other measures are used for quantifying entanglement, average value of distributed entanglement in probabilistic approach is not necessarily equal to the amount of entanglement distributed in deterministic approach.\\
\\
During all steps of EDSS protocol, ancilla qubit $c$ is in separable state with rest of the qubits. More precisely entanglement between partitions $c|a$, $c|b$ and $c|ab$ is zero during the process. As qubit $c$  does not have entanglement with other parts of the system to be disturbed by noise, it may be expected that the protocol can be run successfully even if the communication channel between Alice and Bob is  noisy. However, as we will discuss entanglement between partitions $a|bc$ and partitions $b|ac$ generated during the EDSS protocol, are sensitive to noise. Hence, to consider a more realistic situation and also for better understanding the key features behind the success of EDSS protocol, in what follows we consider the situation ancilla $c$ experience noise when it is transferred from Alice to Bob. 
\section{Quantum Communication Channels}\label{QuantumChannels}
In EDSS protocol distant qubits in Alice's and Bob's lab interact with each other through the exchange particle namely qubit $c$. In ideal scenario, this transmission is done through an ideal communication channel which has no effect on this qubit. This is while in realistic situations, communication channel expose noise on the exchange qubit. Before going to the details on noise effect on EDSS protocol, we remind the reader that any kind of noise on a system with density matrix $\rho$, is described by completely positive trace preserving (CPT) map $\mathcal{E}$:
\begin{equation}
\mathcal{E}(\rho)=\sum_k A_k\rho A_k^{\dag},
\end{equation}
where $A_k$'s are known as Kraus operators and satisfy $\sum_k A_k^{\dag}A_k=I$. 
For the case of CPT maps on qubits, another useful channel representation is given by affine map. To introduced the affine map, characterization of qubit by Bloch vector $\vec{r}$ is used. It is known that any qubit can be represented as follows:
\begin{equation}
\rho=\frac{1}{2}(I+\vec{r}.\sigma)
\end{equation}
where $I$ is two dimensional identity matrix, $\sigma_i$'s are Pauli operators and $\vec{r}$ is Bloch vector with $|r|<1$. In this representation, any CPT map $\mathcal{E}$ is described by affine map as follows:
\begin{equation}
\mathcal{E}: \vec{r}\rightarrow\Lambda \vec{r}+\vec{d}.
\end{equation}
Where $\Lambda$ is a three by three matrix and $\vec{d}$ is a three dimensional vector. In \cite{KingRuskai} it has been shown that by change of basis, any qubit CPT map $\mathcal{E}$ corresponds to a canonical CPT map $\mathcal{E}_c$ described by:
\begin{equation}\label{canonical}
\Lambda_d=\left(
\begin{array}{ccc}
\lambda_1& 0&0\cr
0 &\lambda_2&0\cr
0&0 &\lambda_3
\end{array}
\right),\quad 
t=\left(
\begin{array}{c}
t_1\cr
t_2\cr
t_3
\end{array}
\right).
\end{equation}
That is $\mathcal{E}(\rho)=U\mathcal{E}_{c}(V\rho V^{\dag})U^{\dag}$. Indeed completely positivity impose constraints on the elements of $\Lambda_d$ and $\vec{t}$ \cite{KingRuskai,Fujiwara}. In our analysis, we restrict our attention to canonical channels as characterized in equation (\ref{canonical}) with $t_1=t_2=0$. This class includes important family of quantum channels. For $t_3=0$, it represents the general form of unital channels (channels that map identity to identity) which correspond to the important family of Pauli channels. When $t_3\neq 0$, the channel is non-unital and important amplitude damping channel is included in this subclass. Furthermore, extreme points of the set of CPT maps are included in this subclass. It has been shown that a channel characterized by equation \eqref{canonical}, is an extreme point of the set of CPT maps if and only if at most one of the $t_k$s is non-zero (by convention this is $t_3$) and also $(\lambda_1\pm\lambda_2)^2=(1\pm\lambda_3)^2-t_3^2$ \cite{Ruskai}. The fact that any CPT map can be written in terms of these extreme points, highlights the importance of considering this class of quantum channels for our analysis. Furthermore, this class covers a large family of qubit channels which usually appear in experimental settings. Hence analysing the effect of this class of noisy channels on EDSS protocol provides us with great insight about the robustness of this protocol against wide range of noises. 
\section{Noise effects on distributing two-qubit entangled states} \label{Noise2-qubit}
In this section we study a more realistic scenario for entanglement
distribution where the communication channel is noisy. We consider class of noisy channels characterized by equation \eqref{canonical} with $t_1=t_2=0$ and $t_3=t$. After presenting the general results we discuss two important examples of depolarizing and amplitude damping channels. \\
\\
In EDSS protocol, after the preparation made in Alice's lab, state of three qubits is described by $\r^{(1)}_{abc}$ in equation \eqref{eq:firstcnot}. By sending the ancillary qubit $c$ through the noisy channel state of all qubits is given by
\begin{eqnarray}
\r^{(1)'}_{abc}&=&\mathcal{E}(\r^{(1)}_{abc})\cr
&=&\frac{1}{6}\sum_{m=0}^1\Pi_{m,m}\otimes(I+t\sigma_z)\cr
&+&\frac{1}{12}\sum_{m\neq n}\Pi_{m,n}\otimes\Big(I+(t+(-1)^m\lambda_3)\sigma_z\Big)\cr\cr
&+&\frac{1}{12}\sum_{m\neq n}|m,m\rangle\langle n,n|\otimes(\lambda_1\sigma_x+i(-1)^m\lambda_2\sigma_y)\nn\\
\end{eqnarray}
When Bob receives qubit $c$, he follows the regular steps of EDSS algorithm by applying CNOT on qubits $b$ and $c$. Since 
\begin{equation}\label{C12}
C_{1,2}(|m\rangle_1\langle n|\otimes\rho_2)=|m\rangle_1\langle n|\otimes\sigma_x^m\rho_2\sigma_x^n,
\end{equation}
after action of CNOT by Bob, the state of three qubits is described by
\begin{equation}\label{rho'_abc}
\rho_{abc}^{(2)'}=C_{bc}({\r}_{abc}^{(1)'})=
q_{0}\r^{(0)}_{ab}\otimes\v 0\ra\la 0\v+q_{1}\r^{(1)}_{ab}\otimes\v 1\ra\la 1\v,
\end{equation}
where 
\begin{equation}
q_l=\frac{3-g_l}{6},\quad l=0,1,
\end{equation}
and
\begin{eqnarray}\label{rho_ab}
\r^{(l)}_{ab}=\frac{1}{12q_l}&\Bigg(&2h_l\sum_{m\neq n}|m\rangle\langle n|^{\otimes 2}+2\sum_{m=0}^1f_{l+m}\Pi_{m,m}\cr\cr
&+&\sum_{m}(f_{l+m+1}-g_l)\Pi_{m,m+1}\Bigg),
\end{eqnarray}
in which
\begin{eqnarray}\label{co}
f_{l}&:=&\Big(1+(-1)^{l}t\Big)\cr\cr
g_l&:=&(-1)^l\lambda_3\cr\cr
h_l&:=&\frac{1}{2}(\lambda_1+(-1)^l\lambda_2).
\end{eqnarray}
Hence if Bob, measures qubit $c$ in computational basis, state of qubits $a$ and $b$ is projected either to $\rho_{ab}^{(0)}$ or $\rho_{ab}^{(1)}$ with probability $q_0$ and $q_1$ respectively. Therefore, the average distributed entanglement 
quantified by negativity \cite{neg1} (For details about negativity see Appendix \ref{AppNeg}) is given by
\be\label{Nbar_ab}
\bar{N}(\r_{ab})=q_{0}N_{a|b}(\r^{(0)}_{ab})+q_{1}N_{a|b}(\r^{(1)}_{ab}).
\ee
On the other hand, entanglement between $a$ and $bc$ before Bob's measurement, i.e entanglement in partitions $a|bc$ of $\rho_{abc}^{(2)'}$ is given by
\ba\label{Na|bc}
N_{a|bc}(\r^{(2)'}_{abc})&=&q_{0}N_{a|bc}(\r^{(0)}_{ab}\otimes\v 0\ra\la 0\v)+q_{1}N_{a|bc}(\r^{(1)}_{ab}\otimes\v 1\ra\la 1\v)\nn\\
&=& q_{0}N_{a|b}(\r^{(0)}_{ab})+q_{1}N_{a|b}(\r^{(1)}_{ab})\nn\\
&=&\bar{N}(\rho_{ab}),
\ea
where in the first equality we use the block diagonal form of $\rho_{abc}^{(2)'}$ and second equality is valid because the negative eigenvalues of $X\otimes |k\rangle\langle k|$ are equal to negative eigenvalues of $X$. Last equality is found by comparing equations \eqref{Nbar_ab} and \eqref{Na|bc}. By repeating this argument for entanglement between partitions $b$ and $ac$ we find that
\begin{equation}\label{Nb|ac}
N_{b|ac}(\r^{(2)'}_{abc})=\bar{N}(\rho_{ab}).
\end{equation}
Considering that entanglement between $a$ and $bc$ does not change under unitary evolution of $bc$ we have
\begin{equation}\label{NlocalU}
N_{a|bc}(\rho_{abc}^{(1)'})=N_{a|bc}(\rho_{abc}^{(2)'}).
\end{equation}
By passing qubit $c$ through the noisy channel entanglement in partitions $a|bc$ is disturbed. From equations \eqref{Na|bc}, \eqref{Nb|ac} and \eqref{NlocalU} we conclude that the amount of distillable entanglement remained in partition $a|bc$ after noise affects $c$, is mapped to distillable entanglement between partitions $a|bc$ and $b|ac$ by action of CNOT on qubits $b$ and $c$ in Bob's lab. This amount of distillable entanglement is the average of entanglement one gains between qubits $a$ and $b$ if qubit $c$ is measured in computational basis. In summary \\
\begin{equation}\label{All}
\bar{N}_{a|b}(\rho_{ab})=N_{a|bc}(\rho_{abc}^{(1)'})=N_{a|bc}(\rho_{abc}^{(2)'})=N_{b|ac}(\rho_{abc}^{(2)'}).
\end{equation}
Hence if noise completely destroys the distillable entanglement between partitions $a|bc$ and $b|ac$, average distributed entanglement $\bar{N}_{a|b}(\rho_{ab})$, is zero or in other words, no distillable entanglement is distributed among distant qubits $a$ and $b$. This result is valid for a large class of channels described by \eqref{canonical} with $t_1=t_2=0$ and highlights the fact that the success of EDSS protocol relies on the entanglement generated between partitions $a|bc$ and $b|ac$ in intermediate steps of the protocol.\\
\subsection{Depolarising channel} \label{subsec:depolarising two qubits}
An important subset of quantum channels, are unital channels which are described by affine transformation in \eqref{canonical} with $\vec{t}=0$. All Pauli channels are obtained by proper choice of $\lambda_1$, $\lambda_2$ and $\lambda_3$. One important example of Pauli channels is depolarizing channel in which all Pauli operators perform as error operators with same probability. It is straightforward to show that such a channel is described as follows:
\begin{equation}
\mathcal{E}(\rho)=(1-p)\rho+\frac{p}{2}I
\end{equation}
As it is seen in the above equation, each input state remains invariant with probability $(1-p)$ and it may be changed to a completely mixed state (no information
from initial state is remained in the output) with probability $p$. Hence it is expected that such a communication channel has strong
effects on any protocol including EDSS protocol. This channel corresponds to an affine map of form \eqref{canonical} with $\lambda_1=\lambda_2=\lambda_3=1-p$ and $\vec{t}=0$. Regarding equations \eqref{rho'_abc} and \eqref{rho_ab}, state shared between three qubits $a,b$ and $c$ before performing measurement by Bob is described by 
\be \label{eq2}
\r^{(2)'}_{abc}=q_{0}\r^{(0)}_{ab}\otimes\v 0\ra\la 0\v+(1-q_{0})\r^{(1)}_{ab}\otimes\v 1\ra\la 1\v,
\ee
in which
\be
q_{0}=\frac{2+p}{6},
\ee
and
\ba
\r^{(0)}_{ab}&=&\frac{1}{2+p}\Big(\sum_{m}\Pi_{mm}+\frac{p}{2}\sum_{m\neq n}\Pi_{mn}+(1-p)\sum_{m\neq n}|m\rangle\langle n|^{\otimes 2}\Big),\nn\\
\r^{(1)}_{ab}&=&\frac{1}{4-p}\Big(\sum_{m}\Pi_{mm}+\frac{2-p}{2}\sum_{m\neq n}\Pi_{mn}\Big).
\ea
When Bob measures qubit $c$ in computational basis, 
if the outcome of the measurement is $|1\rangle$ a separable state $\rho^{(1)}_{ab}$  is shared between Alice and Bob and the protocol is unsuccessful in distributing entanglement between qubits $a$ and $b$. But if the outcome of measurement is $|0\rangle$,  state of qubits $a$ and $b$ is projected onto an entangled state $\rho_{ab}^{(0)}$ with success probability $p_s=q_{0}$. In this case the amount of shared entanglement quantified by negativity is given by. 
\be\label{Crhoabqubit}
N_{a|b}(\rho_{ab}^{(0)})=\bigg\lbrace\begin{array}{cc}
\frac{2-3p}{2+p},&0\leq p\leq \frac{2}{3}\cr\cr
0,&\frac{2}{3}\leq p\leq 1
\end{array}.
\ee
Hence average value of shared entanglement is found to be
\begin{equation} \label{ave1}
\bar{N}_{a|b}(\rho_{ab})=p_{s} N_{a|b}(\rho_{ab}^{{(0)}})=
\bigg\lbrace\begin{array}{cc}
\frac{2-3p}{6},&
0\leq p\leq \frac{2}{3}\cr\cr
0,&\frac{2}{3}\leq p\leq 1
\end{array}.
\end{equation}
As expected average of shared entanglement decreases as the noise parameter increases and vanished for $p\geq\frac{2}{3}$. Regarding 
equation \eqref{All}, we conclude that when noise parameter goes beyond this critical value, $p_c=\frac{2}{3}$, distillable entanglement in partitions $a|bc$ and $b|ac$ is broken and hence protocol is not successful in distributing entanglement. 
\\
\\
If instead of measuring qubit $c$, Bob applies the quantum channel in
\eqref{LCPTMAP},
on qubits $b$ and $c$,  the outcome is described by
\begin{eqnarray}
\chi_{ab}&=&\mathrm{tr}_{c}(\Phi_{bc}(\rho_{abc}^{(2)'}))\cr\cr
&=&\frac{1-p}{3}(|\psi^+\rangle\langle\psi^+|+\mathrm{I}\otimes |0\rangle\langle 0|)\cr\cr
&+&\frac{p}{12}(\sum_{i=0}^1 |i,i\rangle\langle i,i|+\mathrm{I}\otimes\mathrm{I}+3\mathrm{I}\otimes |0\rangle\langle 0|).
\end{eqnarray}
In this deterministic approach the amount of shared entanglement between Alice and Bob for $0\leq p<\frac{3-\sqrt{5}}{2}$ is:
\begin{figure}
 \centering
\includegraphics[scale=0.32]{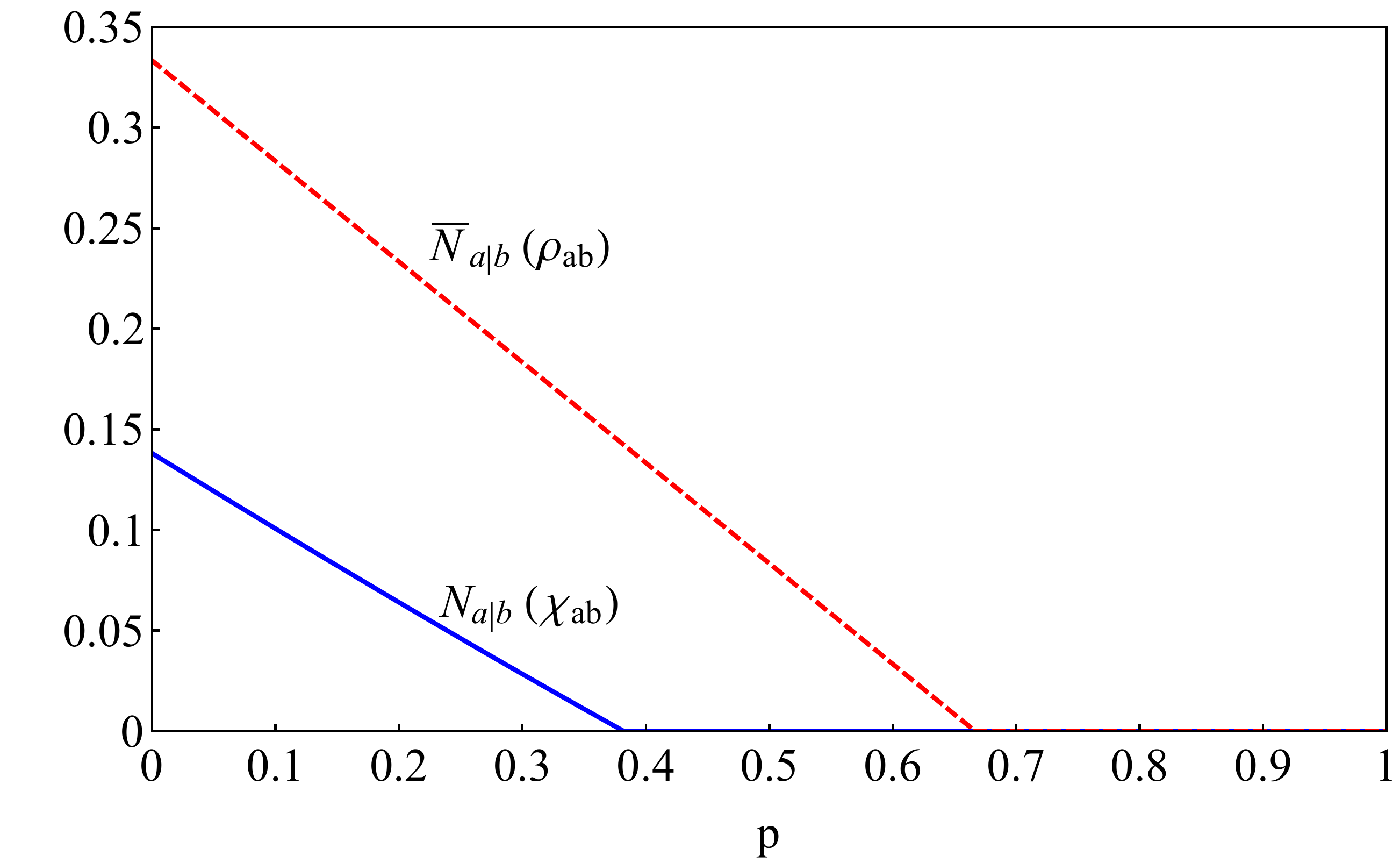}
\caption{Dashed red line: Average entanglement distributed between qubits $a$ and $b$ (equation \eqref{ave1}) in probabilistic approach versus noise parameter $p$. Solid blue line: Entanglement distributed between qubits $a$ and $b$ (equation \eqref{Cs}) in deterministic approach versus noise parameter $p$. In both cases the communication channel in depolarizing channel. }
 \label{fig9}
\end{figure}
\begin{equation}\label{Cs}
N_{a|b}(\chi_{ab})=\frac{1}{12}(\sqrt{17p^2-40p+32}-p-4),
\end{equation}
which is a decreasing function of $p$. For $\frac{3-\sqrt{5}}{2}\leq p\leq1$, entanglement can not be shared between Alice and Bob.
Average value of shared entanglement in probabilistic approach (equation \eqref{ave1}) and value of distributed entanglement in deterministic approach (equation \eqref{Cs}) are shown in figure (\ref{fig9}) versus noise parameter $p$. As it is seen in this figure, in probabilistic approach higher value of entanglement can be distributed on average.  Moreover, probabilistic approach is more robust against noise in the sense that the protocol is successful up to a higher value of noise parameter.
\subsection{Amplitude damping channel}
As an important example of non-unital channels we consider amplitude damping channel which models a typical source of noise resulting from interaction of a single atom with a bosonic bath. This channel is characterized by $\lambda_1=\lambda_2=\sqrt{1-\gamma}$, $\lambda_3=\lambda_1^2$ and  $t_3=\gamma$. Apart from its practical applications, amplitude damping channel has interesting theoretical characteristics such as being an extreme point of the set of CPT maps. Motivated by these, in this subsection we describe the effect of this channel on EDSS protocol.\\
\\
Following the general solution given in equations \eqref{rho'_abc} and \eqref{rho_ab}, state of three qubits after passing qubit $c$ through the amplitude damping channel and performing $ \mathrm{CNOT} $ on qubits b and c is described by:
\be
\r^{(2)'}_{abc}=p_{0}\r^{(0)}_{ab}\otimes\v 0\ra\la 0\v+(1-p_{0})\r^{(1)}_{ab}\otimes\v 1\ra\la 1\v,
\ee
where $p_{0}=\frac{2+\g}{6}$ and
\ba\label{rho_ab_AD}
\r^{(0)}_{ab}&=&\frac{1}{2+\g}\Big(\sum_m(1+(-1)^m\g)\Pi_{m,m}+\g\Pi_{10}\cr\cr
&&+\sqrt{1-\g}\sum_{i\neq j}\v i,i\ra\la j,j\v\Big),\nn\\
\r^{(1)}_{ab}&=&\frac{1}{4-\g}\Big(I\otimes I-\gamma(I\otimes\Pi_0-\Pi_{1,1})\Big).\cr
\ea
When Bob measures qubit $c$, if the outcome of measurement is $|0\rangle$, entangled state $\rho_{ab}^{(0)}$ as in equation \eqref{rho_ab_AD} is shared between qubits $a$ and $b$. Negativity of this state is given by
\be
N_{a|b}(\r^{(0)}_{ab})=\frac{2-2\g}{2+\g},\,\,\,\,\,\,0\leq\g\leq 1.
\ee
Hence the average value of entanglement shared between qubits $a$ and $b$ at distance labs is found to be
\be\label{ave1A}
\bar{N}_{a|b}(\r_{ab})=p_{0}N_{a|b}(\r^{(0)}_{ab})=\frac{1-\g}{3}.
\ee
As expected average of shared entanglement decreases as the noise parameter increases. It is worth noticing that unlike depolarizing noise, when the communication channel is amplitude damping, it is possible to distribute entanglement for all values of noise parameters except $\gamma=1$. In other words depolarizing channels appears to be more destructive for EDSS protocol. 
\\
\\
If instead of measuring qubit $c$, Bob applies the quantum channel in equation \eqref{LCPTMAP} on qubits $b$ and $c$,  the outcome is described by
\begin{eqnarray}
\chi_{ab}&=&\frac{1}{6}\Bigg(\sum_{m}(1+(-1)^m\gamma)\Pi_{mm}+2\gamma\Pi_{10}\cr\cr
&+&(2-\gamma)I\otimes \Pi_0+\sqrt{1-\gamma}\sum_{m\neq n}|mm\rangle\langle nn|
\end{eqnarray}
We can quantify the entanglement of this state by using Negativity:
\be \label{CsA}
N_{a|b}(\chi_{ab})=\frac{\sqrt{8+\g^{2}}-2-\gamma}{6}
\ee
\begin{figure}
 \centering
\includegraphics[scale=0.32]{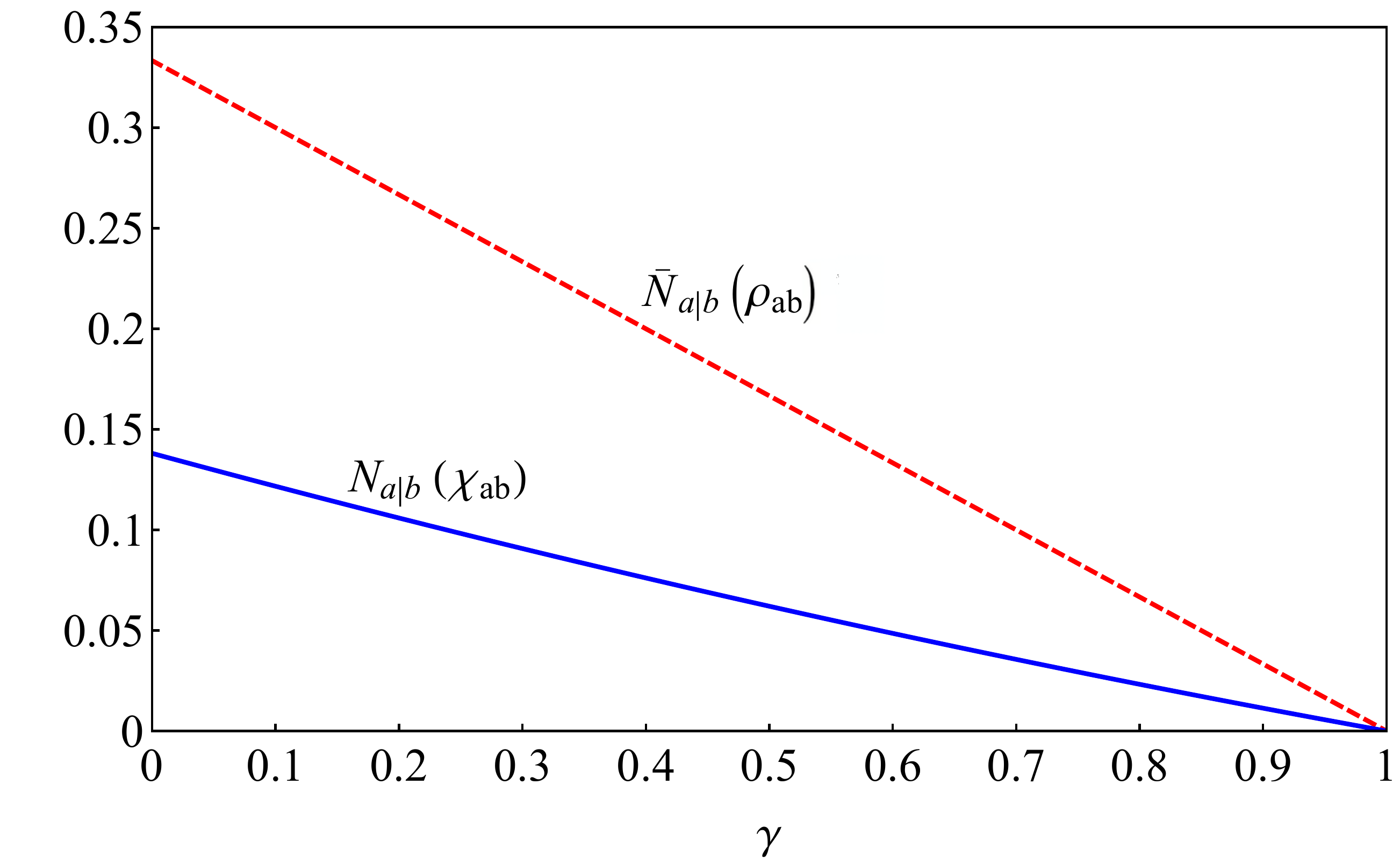} 
\caption{Dashed red line: Average entanglement distributed between qubits a and b (equation \eqref{ave1A}) in probabilistic approach versus noise
parameter $ \g $ . Solid blue line: Entanglement distributed between qubits a and b (equation \eqref{CsA}) in deterministic approach versus noise parameter
$ \g $. In both cases the communication channel is amplitude damping channel.}
\label{fig1}
\end{figure}
Average value of shared entanglement in probabilistic approach (equation \eqref{ave1A}) and value of distributed entanglement in deterministic approach (equation \eqref{CsA}) are shown in figure (\ref{fig1}) versus noise parameter $\g$. As it is seen in this figure, in probabilistic approach higher value of entanglement can be distributed on average. Similar effect is seen in (\ref{fig9}) when communication channel is depolarizing channel. Hence in what follows we focus on the probabilistic approach for
distributing GHZ state in presence of noise. Distribution of d-dimensional maximally entangled states is addressed in appendix \ref{dBell-SEDSS & Noise}.
\section{Noise effects on distributing three-qubit entangled states}\label{General Noise on GHZ}
While in the previous sections we studied the noise effect on distributing entanglement between two parties,
in this section we study the effect of noise on distributing entanglement between qubits $a$, $b$ and $c$ which are respectively in Alice's, Bob's and Charlie's labs. In \cite{sedss} it is shown that by using two ancillary qubits $d_1$ and $d_2$ it is possible to distribute a GHZ state between three distant labs with probabilistic EDSS protocol, if the initial separable state is prepared in the following form 
\ba \label{eq:3GHZ0}
\sigma^{(0)}_{abc\mathcal{D}}&=&\frac{4}{49}\sum_{k=0}^{6}|\omega(k)\ra\la\omega(k)|\otimes\Pi_{0,0}\cr\cr
&+&\frac{1}{14}\sum_{m=0}^{1}\Pi_{mmm}\otimes
(\mathrm{I}\otimes\mathrm{I}-\Pi_{00}),
\ea
in which we have used the abbreviated notation $\mathcal{D}=d_1d_2$. 
and
\begin{equation*}
|\omega(k)\ra_{abc}=\v\phi_{1}(k),\phi_{2}(k),\phi_{3}(k)\ra,
\end{equation*}
with 
\begin{equation}
\v\phi_{n}(k)\ra=\frac{1}{\sqrt{2}}(\v 0\ra+e^{\frac{2^n\pi ik}{7}}\v 1\ra).
\end{equation}
Initially ancillary qubits are in Alice lab. She applies CNOT gates to qubits $a$ and $d_1$ and also on $a$ and $d_2$ ( in both cases qubit $a$ is controlled qubit) which results the following state:
\begin{eqnarray}\label{eq:3GHZ1}
\sigma^{(1)}_{abc\mathcal{D}}&=&\frac{1}{7}\v\mathrm{GHZ}_{5}^{(2)}\ra\la\mathrm{GHZ}_{5}^{(2)}\v\cr\cr
&+&\frac{1}{14}\bigg(\sum_{i=1}^{3}(\Pi_{0\overline{0i}}+\Pi_{0\overline{i0}})+\sum_{i=0}^{2}(\Pi_{1\overline{3i}}+\Pi_{1\overline{i3}})\bigg).\nn\\
\end{eqnarray}
where $ \v\overline{0}\ra\equiv\v 00\ra $, $ \v\overline{1}\ra\equiv\v 01\ra $, $ \v\overline{2}\ra\equiv\v 10\ra $, $ \v\overline{3}\ra\equiv\v 11\ra $. Actually by performing CNOT, Alice generates correlatation between qubit $a$ and ancillary qubits $d_1$ and $d_2$ through which all qubits $a$, $b$ and $c$ must interact with each other. Then, she sends ancillary qubits $ d_{1} $ and $ d_{2} $, respectively to Bob and Charlie through identical independent channels characterized as in equation \eqref{canonical} with $t_1=t_2$=0 and $t_3=t$. Hence after Bob and Charlie receive the ancillary qubits from Alice the whole state is described by 
\begin{eqnarray}
\s^{(1)'}_{abc\mathcal{D}}&=&(\mathcal{E}_{d_1}\otimes\mathcal{E}_{d_2})(\sigma^{(1)}_{abc\mathcal{D}})\nn\\
&=&\frac{1}{14}\sum_{m\neq n}|m\rangle\langle n|^{\otimes 3}\otimes\mathcal{E}(|m\rangle\langle n|)^{\otimes 2}\cr\cr
&+&\frac{1}{14}\sum_m\Pi_{m}^{\otimes 3}\otimes\Big({\mathcal{E}(I)}^{\otimes 2}-{\mathcal{E}(\Pi_m)}^{\otimes 2}\Big)\cr\cr
&+&\frac{1}{14}\sum_m\Pi_{m}\otimes I^{\otimes 2}\otimes {\mathcal{E}(\Pi_m)}^{\otimes 2},\nn\\
\end{eqnarray}
where
\begin{eqnarray}
\mathcal{E}(\Pi_m)&=&\frac{1}{2}\Big(I+(t+(-1)^m\lambda_3)\sigma_z\Big)\cr\cr
\mathcal{E}(|m\rangle\langle n|)&=&\frac{1}{2}\Big(\lambda_1\sigma_x+i(-1)^m\lambda_2\sigma_y\Big)\quad m\neq n
\end{eqnarray}
Then Bob and Charlie perform CNOT gates on qubits $b-d_1$ and $c-d_2$ (ancillary qubits are target qubits) and produce the following state:
\be\label{GHZgen}
\s^{(2)'}_{abc\mathcal{D}}=\sum_{l,l'=0}^{1}q_{_{l,l'}}
\s_{abc}^{(l,l')}\otimes|l,l'\rangle\langle l,l'|,
\ee
where
\begin{equation}
q_{_{l,l'}}=\frac{8+3(1-f_l)(1-f_{l'})-(1+g_l)(1+g_{l'})}{28},
\end{equation}
and
\begin{eqnarray}
\s_{abc}^{(l,l')}&=&\frac{1}{56q_{_{l,l'}}}\Bigg(4h_lh_{l'}\sum_{m\neq n}|m\rangle\langle n|^{\otimes 3}\cr\cr
&+&\sum_{m}\Big(4f_{l+m}f_{l'+m}-(f_{l+m}+g_l)(f_{l'+m}+g_{l'})\Big)\Pi_{m}^{\otimes 3}\cr\cr
&+&\sum_{m,n,n'}(f_{l+n}+g_{l+m+n})(f_{l'+n'}+g_{l'+m+n'})\Pi_{mnn'}\Bigg).\nn\\
\end{eqnarray}
Coefficients $f_l$, $g_l$ and $h_l$ are defined in equations \eqref{co}. If measuring ancillary qubits $d_1$ and $d_2$ in computational basis results $|l.l'\rangle_{d_1,d_2}$,
state $\sigma^{(l,l')}_{abc}$ is shared between qubits $a$ and $b$ and $c$ with probability $q_{_{l,l'}}$. Hence 
\begin{equation}\label{Nave3}
\bar{N}_{x|yz}(\s_{abc})=\sum_{l,l'}q_{_{l,l'}}N_{x|yz}(\s^{(l,l')}_{abc})
\end{equation}
where $x|yz$ can be any permutation of $a|bc$. On the other hand block diagonal structure of density matrix in 
equation \eqref{GHZgen} suggests that the 
\begin{eqnarray}\label{Nx|yzD}
N_{x|yz\mathcal{D}}(\s^{(2)'}_{abc\mathcal{D}})&=&\sum_{l,l'}q_{_{l,l'}}N_{x|yz\mathcal{D}}(\s^{(l,l')}_{abc}\otimes |l,l'\ra\la  l,l'|)\cr\cr
&=&\sum_{l,l'}q_{_{l,l'}}N_{x|yz}(\s^{(l,l')}_{abc})
\end{eqnarray}
Comparing equations \eqref{Nave3} and \eqref{Nx|yzD} we conclude that the average value of distillable entanglement shared in partition $x|yz$ is equal to the entanglement in partition $x|yz\mathcal{D}$ before the measurement. \begin{equation}
\bar{N}_{x|yz}(\s_{abc})=N_{x|yz\mathcal{D}}(\s^{(2)'}_{abc\mathcal{D}}).
\end{equation}
Furthermore, since local unitary operation do not change the value of negativity we have: 
\begin{equation}\label{GHZGench}
\bar{N}_{a|bc}(\s_{abc})={N}_{a|bc\mathcal{D}}(\s^{(2)'}_{abc\mathcal{D}})={N}_{a|bc\mathcal{D}}(\s^{(1)'}_{abc\mathcal{D}}),\\
\end{equation}
which means that the average entanglement distributed in partition $a|bc$ is equal to the amount of entanglement remained in partition $a|bc\mathcal{D}$ after the effect of noise in communication channel. 
\subsection{Depolarising Channel}
In this subsection as an example we assume that the communication channels are depolarising channels, that is $\lambda_1=\lambda_2=\lambda_3=1-p$ and $\vec{t}=0$. Since for this channel $h_l=(1-p)\delta_{l,0}$, (see equation \eqref{co}) it is apparent that 
when Bob and Charlie measure ancilla qubits $d_1$ and $d_2$ in computational basis, if the outcome of measurement is $|00\rangle$, entanglement is distributed between qubits $a$, $b$ and $c$ and state of three qubits is given by
\ba \label{eq:sabcnoise}
\s_{abc}^{(00)}&=&\frac{1}{4+4p-p^{2}}( 4(1-p)^{2}|\mathrm{GHZ_{3}}^{(2)}\ra\la \mathrm{GHZ_{3}}^{(2)}|\cr\cr
&+&\frac{8p-5p^2}{2}\sum_m\Pi_{mmm}\cr\cr
&+&p(1-p) \mathrm{I}\otimes \sum_{m\neq n}\Pi_{mn}+\frac{p^2}{2}\mathrm{I}^{\otimes 3}),
\ea
For other measurement outcomes, state of qubits $a$, $b$ and $c$ is separable, hence the success probability of distributing an entangled states between target qubits is equal to the probability of having $|00\rangle$ when measuring qubits $d_1$ and $d_2$, that is
\be \label{psucc3} 
p_{s}=q_{_{0,0}}=\frac{4+4p-p^{2}}{28},
\ee
In the ideal case that the communication channel is not noisy ($p=0$) if the outcome of measurement is $|00\rangle$, three qubit GHZ state is distributed in which entanglement between each two pairs is zero and each qubit is in maximally entangled state with other two qubits. In presence of noise, in the best case we obtain state $\s_{abc}^{(00)}$ as in equation \eqref{eq:sabcnoise}. It is easy to see that entanglement between each two qubits in $\s_{abc}^{(00)}$ is zero, similar to the ideal case. To analyse entanglement between one of the qubits and the rest of the system we compute negativity in different partitions:
\\
\begin{equation}\label{N3parts_a}
N_{a|bc}(\sigma_{abc}^{(00)})=\bigg\lbrace\begin{array}{cc}
\frac{4-8p+3p^2}{4+4p-p^2},& 0\leq p\leq \frac{2}{3}\cr\cr
0,& \frac{2}{3}\leq p\leq 1
\end{array},
\end{equation}
and
\begin{equation}\label{N3parts_b}
N_{b|ac}(\sigma_{abc}^{(00)})=\bigg\lbrace\begin{array}{cc}
\frac{4-10p+5p^2}{4+4p-p^2},& 0\leq p\leq\frac{\sqrt{5}-1}{\sqrt{5}}\cr\cr
0,& \frac{\sqrt{5}-1}{\sqrt{5}}\leq p\leq1
\end{array}.
\end{equation}
Due to the symmetric role of qubits $b$ and $c$ in the protocol we have $N_{b|ac}(\sigma_{abc}^{(00)})=N_{c|ab}(\sigma_{abc}^{(00)})$ and we expect that $N_{a|bc}(\sigma_{abc}^{(00)})$ be different from those because of the different role of qubit $a$ in comparison with qubits $b$ and $c$. Following the general discussions made in this section, we have
\begin{equation}
N_{a|bc\mathcal{D}}(\s^{(2)'}_{abc\mathcal{D}})=
\bar{N}_{a|bc}(\sigma_{abc})=
\bigg\lbrace\begin{array}{cc}
\frac{4-8p+3p^2}{28}, & p\leq\frac{2}{3}\cr
0,&p>\frac{2}{3}
\end{array},
\end{equation}
and
\begin{figure}
\label{fig14}
 \centering
\includegraphics[scale=0.32]{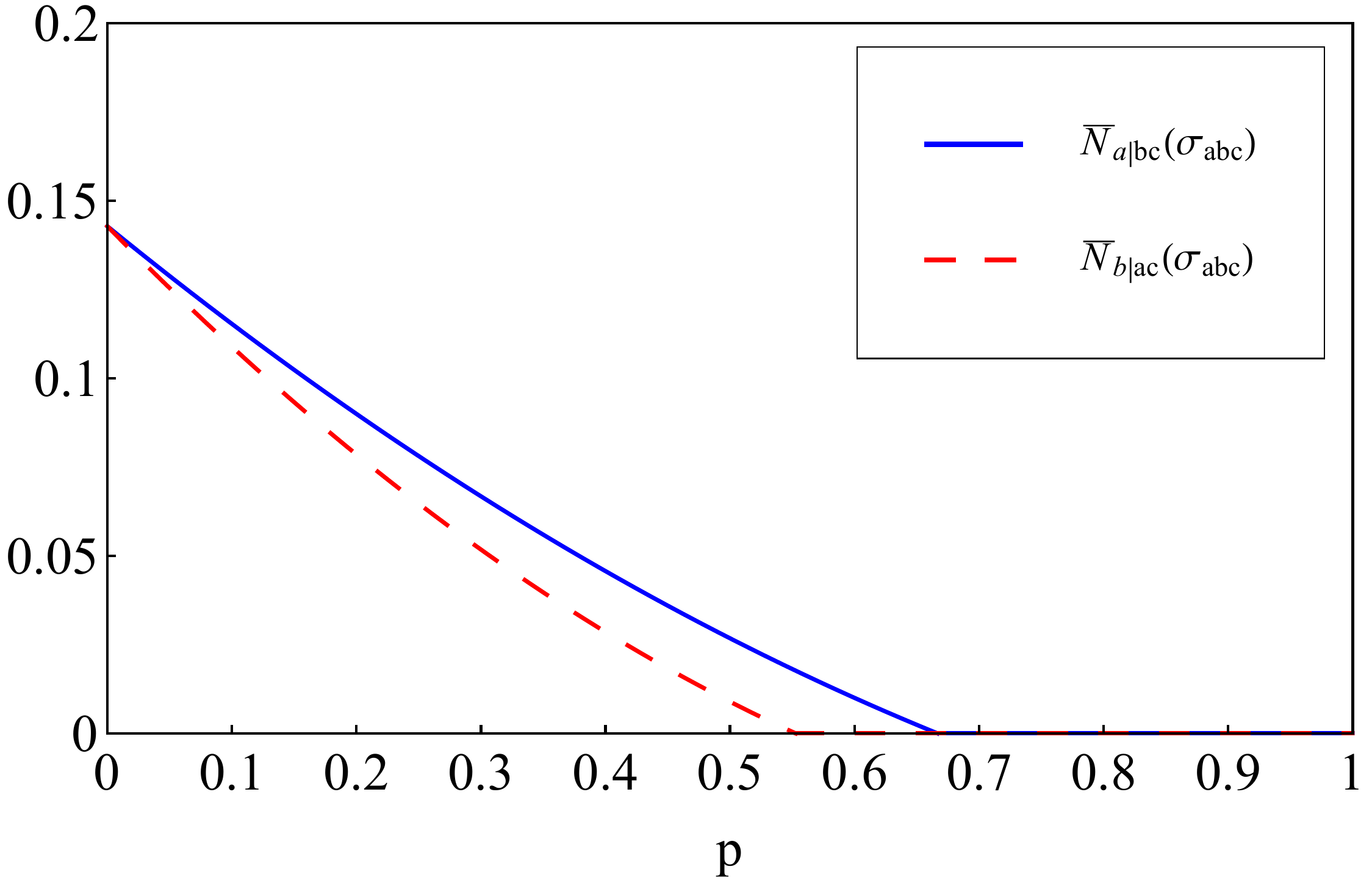}
\caption{Average negativity $\bar{N}_{a|bc}$ (solid blue line) and $\bar{N}_{b|ac}$ (dashed red line) of state generated between qubits $a$, $b$ and $c$ versus noise parameter $ p $ when communication channels are depolarizing channels.}
 \label{fig14}
\end{figure}
\begin{equation}
N_{b|ac\mathcal{D}}(\s^{(2)'}_{abc\mathcal{D}})=\bar{N}_{b|ac}(\sigma_{abc})=
\bigg\lbrace\begin{array}{cc}
\frac{4-10p+5p^2}{28},& p\leq\frac{\sqrt{5}-1}{\sqrt{5}}\cr
0,&p>\frac{\sqrt{5}-1}{\sqrt{5}}
\end{array}.
\end{equation}
Hence when distillable entanglement between $a$ and $bc\mathcal{D}$ breaks down due to the noise effect, no distillable entanglement can be distributed between $a$ and $bc$. Similarly,
if distillable entanglement between $b$ and $ac\mathcal{D}$ vanished due to the noise, no distillable entanglement is shared in partitions $b|ac$ and $c|ab$ ($N_{c|ab\mathcal{D}}(\s^{(2)'}_{abc\mathcal{D}})=N_{b|ac\mathcal{D}}(\s^{(2)'}_{abc\mathcal{D}})=\bar{N}_{b|ac}(\sigma_{abc})$). 
Figure (\ref{fig14}) shows these quantities versus noise parameter $p$. As $p$ increases entanglement in all mentioned bipartition decreases. Furthermore, by increasing $p$, it is seen that $N_{a|bc}(\sigma_{abc})$ (solid blue line) deviates more from $N_{b|ac}(\sigma_{abc})=N_{c|ab}(\sigma_{abc})$  (dashed red line). It shows that the entanglement pattern in $\sigma_{abc}$ deviates more from entanglement pattern of an ideal GHZ state as  $p$ increases. \\
\subsection{Amplitude damping Channel}
\begin{figure}
 \centering
\includegraphics[scale=0.32]{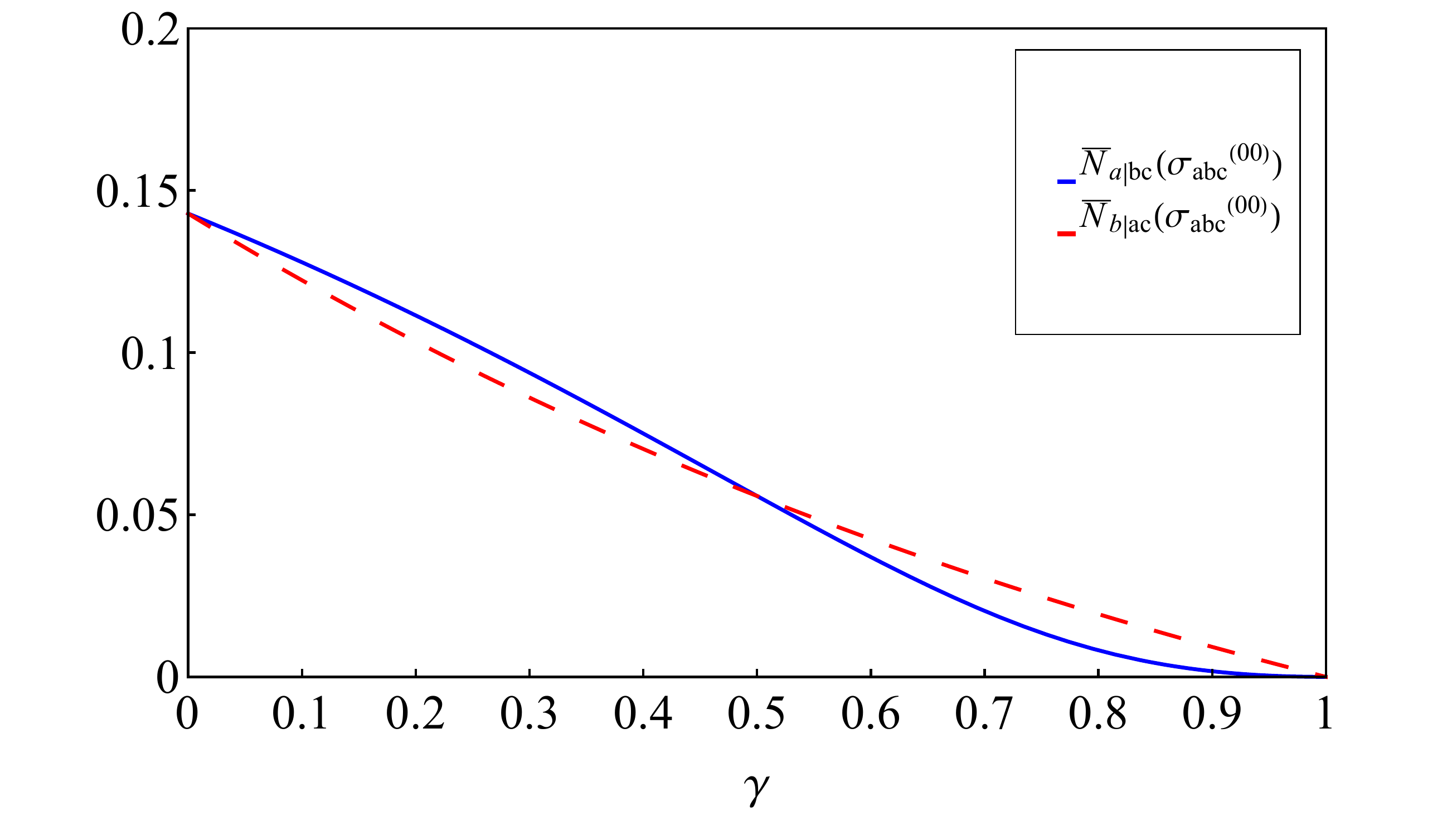} 
\caption{Average negativity $ \bar{N}_{a|bc}$  (solid blue line) and$ \bar{N}_{b|ac}$  (dashed red line) of state generated between qubits a, b and c versus noise parameter $ \g $ when communication channels are amplitude damping channels}
\label{fig3}
\end{figure}

For the case of having amplitude damping channel ($\lambda_1=\lambda_2=\sqrt{1-\gamma}$, $\lambda_3=\lambda_1^2$ and $t_3=\gamma$)
when Bob and Charlie measure ancillary qubits $d_1$ and $d_2$ in computational basis, for all values of measurement separable state is shared between qubits $a$, $b$ and $c$ unless the outcome of measurement is $|00\rangle$. In such a case shared state is given by
\ba
\s^{(00)}_{abc}&=&\frac{1}{2+2\g+\g^{2}}\Big((1-\g)\sum_{m\neq n}|m\rangle\langle n|^{\otimes 3}\nn\\
&+&\sum_m(1+(-1)^m\gamma)^2\Pi_{mmm}\cr\cr
&+&\gamma(1-\gamma)\sum_{m}\Pi_{1,m,m+1}+\gamma^2\Pi_{100}\Big)\nn\\
\ea
with success probability:
\be
p_{s}=q_{00}=\frac{ 2+2\g+\g^{2} }{14},
\ee
Entanglement between different bipartitions of this state are described as:
\be
N_{a|bc}(\s^{(00)}_{abc})=\frac{\sqrt{\gamma^4+(2\gamma-2)^2}-\g^{2}}{\gamma^2+2\g+2},
\ee
and
\be
N_{b|ac}(\s^{(00)}_{abc})=N_{c|ab}(\s^{(00)}_{abc})=\frac{(1-\g)(\sqrt{\g^{2}+4}-\gamma)}{\gamma^2+2\g+2},
\ee
For entanglement between target qubits and exchange qubits before the final measurements we have:
\be
N_{a|bc\mathcal{D}}(\s^{(2)'}_{abc\mathcal{D}})=\bar{N}_{a|bc}(\s^{(00)}_{abc}) =\frac{\sqrt{\gamma^4+4(1-\gamma)^2}-\g^{2}}{14},
\ee
and
\begin{eqnarray}
N_{b|ac\mathcal{D}}(\s^{(2)'}_{abc\mathcal{D}})&=&N_{c|ab\mathcal{D}}(\s^{(2)'}_{abc\mathcal{D}})=\bar{N}_{b|ac}(\s^{(00)}_{abc})\cr\cr
&=&\frac{(1-\g)(\sqrt{\g^{2}+4}-\gamma)}{14}.
\end{eqnarray}
Figure (\ref{fig3}) shows average distillable entanglement distributed between partitions $a|bc$, $b|ac$ and $c|ab$ in presence of amplitude damping noise. As it is seen in this figure, amount of distributed entanglement reduces with noise parameter $\gamma$. Comparison with the case of having depolarizing noise results that the EDSS for distributing three-qubit entangled state is more successful in presence of amplitude damping noise rather than depolarizing noise.  
\section{Summary and Conclusion} \label{Summary and Conclusion}
Consideration of noise effects on EDSS protocol is essential due to its application in realization and expansions of quantum networks. In this work we have shown that there exist an interesting relation between the success of EDSS protocol in presence of noise and robustness of bipartite entanglement in particular partitions of the system, against noise.\\ 
\\
For distributing entanglement between two qubits $a$ and $b$, by means of a separable exchange qubit $c$, we showed that average value of distributed entanglement between $a$ and $b$, is equal to the distillable entanglement in bipartite partitions $a|bc$ and $c|ab$ after ancilla $c$ is sent through a noisy channel of kind \eqref{canonical} with $t_1=t_2=0$. After obtaining the results for this class of noisy channels, we studied depolarizing channel and amplitude damping channel as important examples of this class of quantum channel. By focusing on depolarizing channel in which all errors are equally probable, we have shown that there is a critical value of noise parameter $p_c$, beyond which entanglement between $a|bc$ and $b|ac$ disappear due to noise and hence EDSS protocol is unsuccessful. \\
\\
We showed that the relation found between average value of distributed entanglement between target qubits and entanglement in different bipartite partitions in intermediate steps of the protocol, is also valid when distributing entanglement between three qubits is required. Actually we have shown that for distributing tripartite entangled state between qubits $a$, $b$ and $c$, distillable entanglement between bipartite partitions $a|bc\mathcal{D}$, $b|ac\mathcal{D}$ and $c|ab\mathcal{D}$ (by $\mathcal{D}=d_1d_2$ we denote ancillary qubits) after sending ancillary qubits through the noisy channel, is equal to the average distillable entanglement distributed between $a|bc$, $b|ac$ and $c|ab$, respectively. Depolarizing and amplitude damping channels are discussed as examples of the class on noises studied.\\
\\
In all of our analysis for distributing entanglement between qubits we consider a large and important class of noisy channels. Indeed using the characterization based on affine transformation given in equation \eqref{canonical} plays an important role in obtaining the result. For distributing $d$-dimensional two partite entangled states we restricted our attention to depolarizing and amplitude damping noisy channels. In appendix \ref{dBell-SEDSS & Noise} we have shown that for these two examples even in d-dimensional case average value of distributed entanglement between $a$ and $b$ is equal to distillable entanglement in bipartite partitions $a|bc$ and $c|ab$ after exchange qudit $c$ experiences noise in communication channel. Our studies can be extended in many directions. For example analysing noise effects on EDSS protocol for distributing continuous-variable entangled states or in distributing n-partite GHZ state, it is interesting to see how the performance of the protocol scales with number of the parties in presence of noise in communication channels.
\acknowledgments
We acknowledge financial support by Sharif University of Technology's Office of Vice President for Research under Grant No.G950223. L.M acknowledges hospitality of the Abdus Salam International Centre for Theoretical Physics (ICTP) where parts of this work were completed.

\appendix
\section{Concurrence}\label{concurrenceDef}
Concurrence which is a measure for quantifying  entanglement in a two qubit system describing by density matrix $\rho$ is defined as follows \cite{concurrence}:
\be
C(\r)=\max \lbrace 0,\lambda_{1}-\lambda_{2}-\lambda_{3}-\lambda_{4}\rbrace,\quad \lambda_{1}\geq\lambda_{2}\geq\lambda_{3}\geq\lambda_{4},
\ee
where $ \lambda_{i} $ sorting in decreasing order, are square root of eigenvalues of matrix $\rho\tilde{\rho}$ where $ \tilde{\r} $ is defined as:
\be
\tilde{\r}=(\s_{y}\otimes\s_{y})\r^{*}(\s_{y}\otimes\s_{y}),
\ee
in which $ \r^{*} $ is the complex conjugate of $ \r $ in  computational basis: $\{ \v 00\ra $, $ \v 01\ra $, $ \v 10\ra, \v 11\ra\} $ and $\sigma_y$ is the Pauli matrix: $\sigma_y=-i(|0\rangle\langle 1|-|1\rangle\langle 0|)$.
\section{Negativity}\label{AppNeg}
Negativity is an entanglement measure which is based on an  partial transposition criterion for separability \cite{Press}. For a bipartite system describing by density matrix $\rho\in\mathcal{H}_A\otimes\mathcal{H_{B}}$, it is defined as follows \cite{neg1}:
\be 
N(\r)=\frac{\Vert\r^{T_{A}}\Vert_{1}-1}{d-1},
\ee
where $ \r^{T_{A}} $ is partial transpose of density matrix $ \r $ with respect to partition $A$,  $d=min\{dim\mathcal{H}_A,dim\mathcal{H}_B\}$ and $ \Vert X \Vert =tr\sqrt{X^{\dag}X}$ is the trace norm. Denoting the eigenvalues of $\rho^{T_A}$ by $\lambda_i$s, negativity is given by
\be
N(\r)=\frac{(\sum_{i}|\lambda_{i}|)-1}{d-1},
\ee
It is easy to see that negativity can be written in terms of negative eigenvalues of $\rho^{T_A}$ as follows:
\begin{equation}
N(\rho)=\frac{2\sum_i'|\lambda_i|}{d-1}.
\end{equation}
where the summation $\sum_i'$ is over negative eigenvalues of $\rho^{T_A}$.
\section{Noise effects on distributing two qudit entangled states} \label{dBell-SEDSS & Noise}
In this appendix we analyse the effect of noise on distributing entanglement between two qudits. We consider two types of noise: depolarizing channel and amplitude damping channel.
\subsection{Depolarising channel}
In sub-section \ref{subsec:depolarising two qubits} by analysing the effect of depolarizing channel on EDSS protocol for distributing entanglement between two qubits, we showed that there is a critical value of noise parameter, beyond which entanglement distribution is impossible. It naturally raises some question like how this critical value may depend on dimension of system and whether or not in higher dimensions the protocol performs as well as it does in two dimensional case. To answer these questions, we start our analysis by considering a separable initial state which is shown to be suitable for distributing $d$-dimensional Bell states between Alice and Bob in ideal case \cite{sedss}:
\ba \label{eq:dbell0}
\Omega^{(0)}_{abc}&=&\frac{d}{D(2d-1)}\sum_{k=0}^{D-1}\v\phi(k),\phi(-k),0\ra\la \phi(k),\phi(-k),0\v\cr\cr
&+&\frac{1}{d(2d-1)}\sum_{j\neq l}\Pi_{jj,l-j},
\ea
where
\begin{equation*}
\v\phi(\pm k)\ra=\frac{1}{\sqrt{d}}\sum_{j=0}^{d-1}w^{\pm s_{j}k}\v j\ra,\quad  k=0,1,\ldots,D-1,
\end{equation*}
with $ w=e^{\frac{2\pi i}{D}} $, $ D=2^{d}-1 $ and $ s_{i}=2^{i}-1 $. Alice generates entanglement between $a$ and $bc$ by performing CNOT gate on qudits $a$ and $c$ which are initially in her lab:
\begin{eqnarray} \label{eq:dbell1}
\Omega^{(1)}_{abc}&=&C_{ac}(\Omega^{(0)}_{abc})\cr\cr
&=&\frac{1}{2d-1}\v\mathrm{GHZ_3^{(d)}}\ra\la\mathrm{GHZ}_3^{(d)}\v+\frac{1}{d(2d-1)}\sum_{j\neq l}(\Pi_{jjl}+\Pi_{jlj}),\nn\\
\end{eqnarray}
Exchange qudit $c$, through which qudits $a$ and $b$ interact, is sent to Bob through a depolarizing channel which is defined as follows:
\begin{equation}
\mathcal{E}(X)=(1-p)X+\frac{p}{d} tr(X)\mathrm{I_d},
\end{equation}
where $\mathrm{I}_d$ is $d$-dimensional identity operator. After sending qudit $c$ through the noisy channel to Bob, the state of the all three qudits is described by:
\begin{eqnarray}
\Omega^{(1)'}_{abc}&=&\mathcal{E}(\Omega^{(1)}_{abc})=(1-p)\Omega^{(1)}_{abc}\cr\cr
&+&\frac{p}{d^2(2d-1)}\left(d\sum_{j=0}^{d-1}\Pi_{jj}+\sum_{i\neq j}\Pi_{ij}\right)\otimes \mathrm{I}_d.\cr
\end{eqnarray}
In the next step Bob performs inverse CNOT gate on qudits $b$ and $c$ which gives
\begin{eqnarray}\label{Omega2p}
\Omega^{(2)'}_{abc}&=&C^{-1}_{bc}(\Omega^{(1)'}_{ abc})=(1-p)\Omega^{(2)}_{abc}\cr\cr
&+&\frac{p}{d^{2}(2d-1)}\bigg( (d-1)\sum_{j,k=0}^{d-1}\Pi_{j,j,k-j}+\sum_{ j, k,l=0}^{d-1}\Pi_{j,l,k-l}\bigg),\cr\nn
\\
\end{eqnarray}
with 
\begin{eqnarray} \label{eq:dbell2}
\Omega^{(2)}_{abc}&=&\frac{1}{2d-1}\v\chi_{0}\ra\la\chi_{0}\v\otimes\v 0\ra\la 0\v\cr\cr
&+&\frac{1}{d(2d-1)}\sum_{j\neq l}(\Pi_{j,l,j-l}+\Pi_{j,j,l-j}).
\end{eqnarray}
where $ \v\chi_{0}\ra=\frac{1}{\sqrt{d}}\sum_{j=0}^{d-1}\v jj\ra$  is $d$-dimensional maximally entangled state.
When Bob measures ancilla in computational basis, the state of qudits $a$ and $b$, is projected to separable state for any outcome of measurement by except $|0\rangle$. If the outcome of the measurement is $|0\rangle$, state of qudits $a$ and $b$ is projected to an entangled state $\Omega_{ab}^{(0)} $: 
\begin{figure}
 \centering
\includegraphics[scale=0.32]{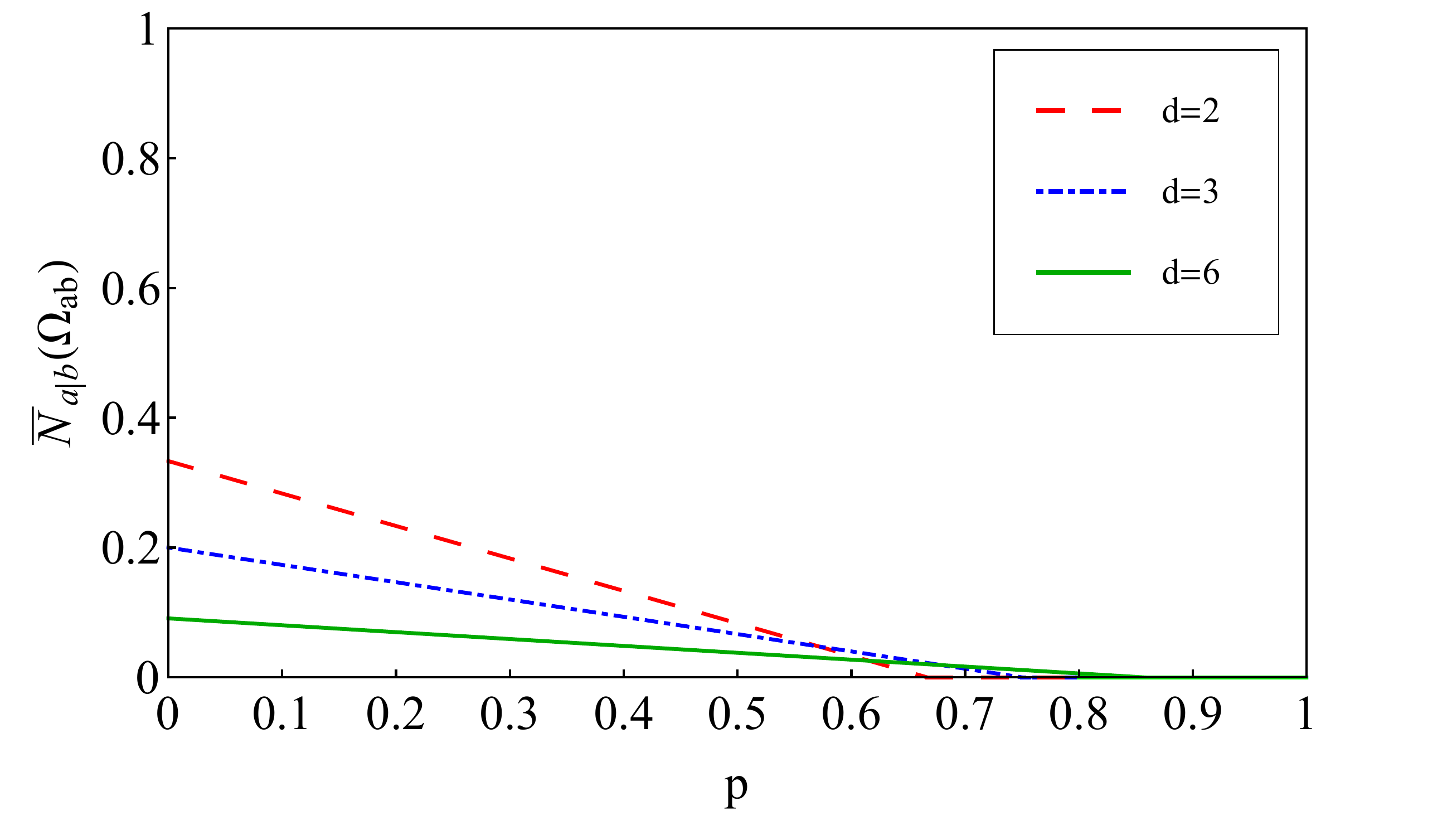}
\caption{Average entanglement shared between qudits $a$ and $b$, (equation \eqref{ave2}) versus noise parameter $p$ for $d=2$ (dashed red line), $d=3$ (dot dashed blue line) and $d=6$ (solid green line) when communication channel is depolarizing channel.}
 \label{fig12}
\end{figure}
\be
\Omega_{ab}^{(0)}=\frac{1}{d+p(d-1)}\bigg(d(1-p)\v\chi_{0}\ra\la\chi_{0}\v+p\sum_{j=0}^{d-1}\Pi_{jj}+\frac{p}{d}\sum_{j\neq l}\Pi_{jl}\bigg),
\ee
Hence success probability of protocol in distributing entanglement between $a$ and $b$ is equal to the probability of having outcome $|0\rangle$ in measuring qudit $c$ and is given by
probability:
\be \label{psucc2} 
p_s=\frac{d+p(d-1)}{d(2d-1)}.
\ee
Quantifying the entanglement properties of $\Omega_{ab}^{(0)}$ by negativity we find that there is a critical value of noise probability 
\begin{equation} \label{eq:pth}
p_{c}=\frac{d}{d+1},
\end{equation}
beyond which negativity is equal to zero and no distillable entanglement can be shared between distant qudits $a$ and $b$:
\begin{equation}\label{NOmegaab}
N(\Omega_{ab}^{(0)})=\Bigg\lbrace\begin{array}{cc}
\frac{d-(d+1)p}{d+(d-1)p},&, 0\leq p\leq p_c\cr\cr
0,&p_c\leq p \leq 1
\end{array}.
\end{equation}
Hence for $p\leq p_c$ average of distillable entanglement between $a$ and $b$ is turned out to be
\begin{equation}\label{ave2}
\bar{N}_{a|b}(\Omega_{ab})=p_sN_{a|b}(\Omega_{ab}^{(0)})=\frac{d-(d+1)p}{d(2d-1)},
\end{equation}
and for $p>p_c$ this quantity is zero. It is worth noticing that state of three qudits before the final measurement has block diagonal form, that is
\begin{equation}
\Omega^{(2)'}_{abc}=p_s\Omega_{ab}^{(0)}\otimes |0\rangle\langle 0|+\sum_{i=1}^{d-1}p_i\Omega_{ab}^{(i)}\otimes |i\rangle\langle i|,
\end{equation}
where $\Omega_{ab}^{(i)}$ for $i=1\cdots d-1$ are separable states. Regarding this block-diagonal form of $\Omega^{(2)'}_{abc}$ and following the same arguments as in section \ref{Noise2-qubit} we conclude that 
\begin{equation}
N_{a|bc}(\Omega^{(2)'}_{abc})=N_{b|ac}(\Omega^{(2)'}_{abc})=\bar{N}_{a|b}(\Omega_{ab}),
\end{equation}
where the first equality is due to the fact that $\Omega^{(2)'}_{abc}$ in equation \eqref{Omega2p} is invariant under permutation of indices $a$ and $b$. Furthermore, 
 since unitary action on qudits $b$ and $c$ can not change the entanglement between partitions $a$ and $bc$ we have
\begin{equation}
N_{a|bc}(\Omega^{(1)'}_{abc})=N_{b|ac}(\Omega^{(2)'}_{abc}).
\end{equation}
Hence what we found for distributing two qubit entangled state is valid for arbitrary dimension. That is, while exchange particle is always in separable state with rest of the system, as long as distillable entanglement between partitions $a|bc$ and $b|ac$ is not vanishing due to the noise, it is possible to distribute distillable entanglement between distant qudits by probabilistic EDSS protocol. 
Figure (\ref{fig12}) shows average distillable entanglement $\bar{N}_{a|b}(\Omega_{ab})$ shared between qudits $a$ and $b$ versus noise 
parameter $p$ for $d=2$ (red dashed line), $d=3$ (blue dash-dotted line) and $d=6$ (green solid line). As it is seen in this figure, by increasing the dimension of Hilbert space, the entanglement decreases more slowly with $ p $. It means that as the dimension increases the protocol is useful for distributing entanglement up to higher value of noise parameter which is given by $p_c$. Figure (\ref{pth}), shows the increase of $p_c$ versus $d$, dimension of Hilbert space.
\begin{figure}
 \centering
\includegraphics[scale=0.30]{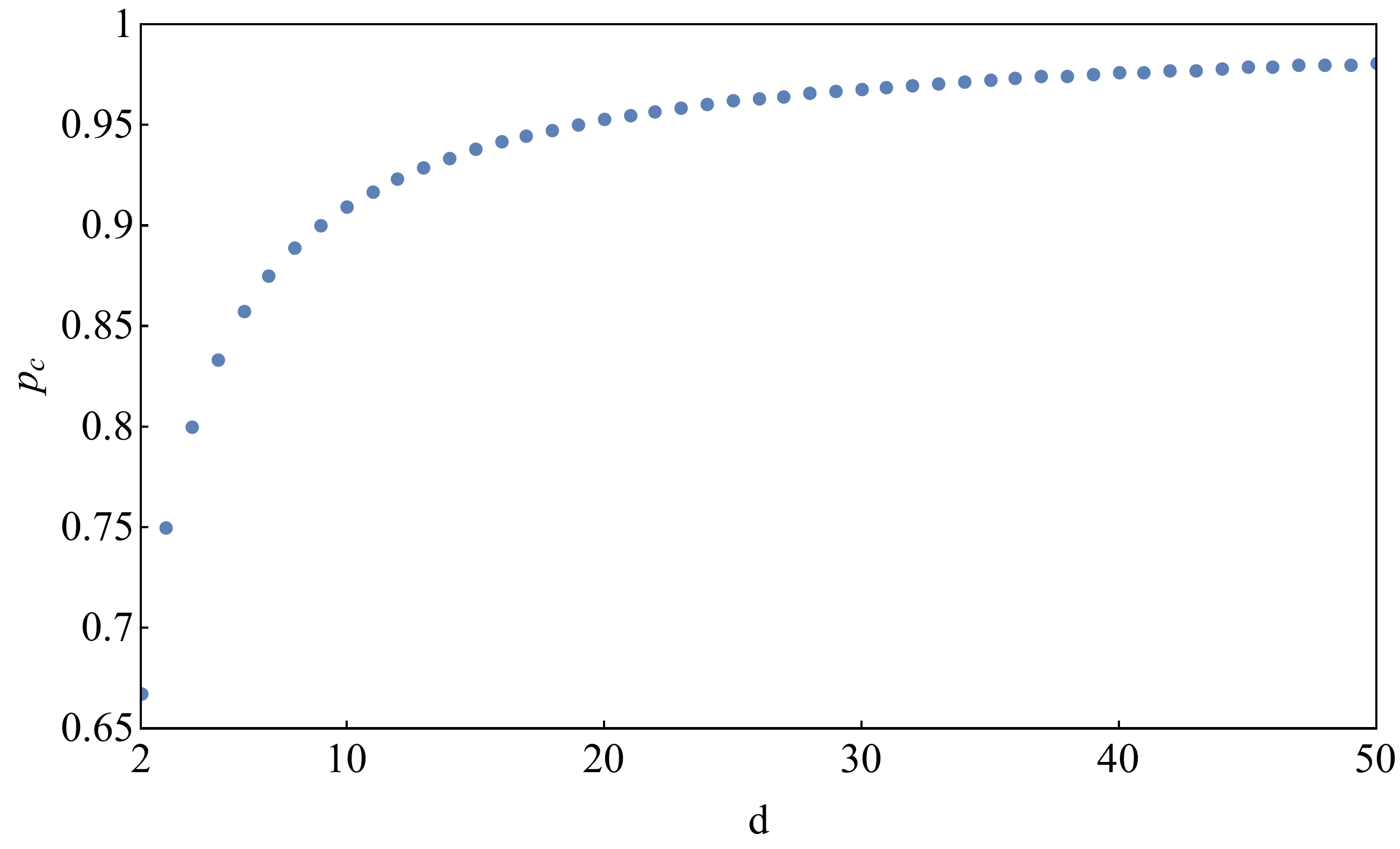}
\caption{Critical value of error parameter $p_{c}$ versus $d$ (dimension of Hilbert space), beyond which distillable entanglement can not be distributed between distant qudits $a$ and $b$ by mediating particle $c$ when communication channel is depolarizing channel.}
 \label{pth}
\end{figure}
\subsection{ Amplitude damping noise}
This part is devoted to analyse the effect of amplitude damping noise on d-dimensional EDSS protocol. Amplitude damping channel on qudits is defined by
\be
\mathcal{E}(\r)=\sum_{m=0}^{d-1}E_{m}\r E_{m}^{\dagger},
\ee
in which
\begin{eqnarray}
E_{0}&=&\v 0\ra\la 0\v+\sum_{i=1}^{d-1}\sqrt{1-\g}\v i\ra\la i|\cr\cr
E_{m}&=&\sqrt{\g}\v 0\ra\la m\v,\quad 1\leq m\leq d-1.
\end{eqnarray}
By applying amplitude damping noise on qudit $c$ of state in equation \eqref{eq:dbell1} we have:
\begin{eqnarray}
\Omega^{(1)'}_{abc}&=&\mathcal{E}_{c}(\Omega^{(1)}_{abc})\nn\\
&=&\frac{1}{d(2d-1)}\Bigg(\gamma(d-1)\Pi_{000}+\Pi_0\otimes I\otimes \Pi_0\cr\cr
&+&\sum_{m=1}\Big(\sqrt{1-\gamma}(|0\rangle\langle m|^{\otimes 3}+|m\rangle\langle 0|^{\otimes 3})\cr\cr
&+&{(1+\gamma(d-1))}\Pi_{mm0}\Big)+\sum_{m,n=1}(1-\gamma)\v m\ra\la n\v^{\otimes 3}\cr\cr
&+&\sum_{m\neq n}\sum_{n=1}\Big((1-\gamma)(\Pi_{mmn}+\Pi_{nmn})+\gamma\Pi_{nm0}\Big)\Bigg)\nn\\
\end{eqnarray}
After applying inverse of $ \mathrm{CNOT} $ on qudits $b$ and $c$, state of three qudits is as follows:
\be\label{D-ADOmega2'}
\Omega^{(2)'}_{abc}=\sum_{m=0}^{d-1}p_{m}\Omega^{(m)}_{ab}\otimes|m\rangle\langle m|,
\ee
It is straightforward to show that when Bob measures qudit $c$ in computational basis, if outcome is $|0\rangle$ shared state between $a$ and $b$ is entangled otherwise it is separable. Therefore by probability 
\begin{equation}
p_{0}=\frac{d+(d-1)\g}{d(2d-1)},
\end{equation}
entangled state 
\ba
\Omega^{(0)}_{ab}&=&\frac{1}{d+(d-1)\g}\Bigg(\big(1+\gamma(d-1)\big)\Pi_{00}\cr\cr
&&\quad+\sum_{m,n=1}(1-\g)\v m\ra\la n\v^{\otimes 2}\cr\cr
&&\quad+\sum_{m=1}^{d-1}\Big(\g\Pi_{m0}+\sqrt{1-\g}(|0\rangle\langle m|^{\otimes 2}+|m\rangle\langle 0|^{\otimes 2})\Big)\Bigg)\nn\\
\ea
\begin{figure}
 \centering
\includegraphics[scale=0.32]{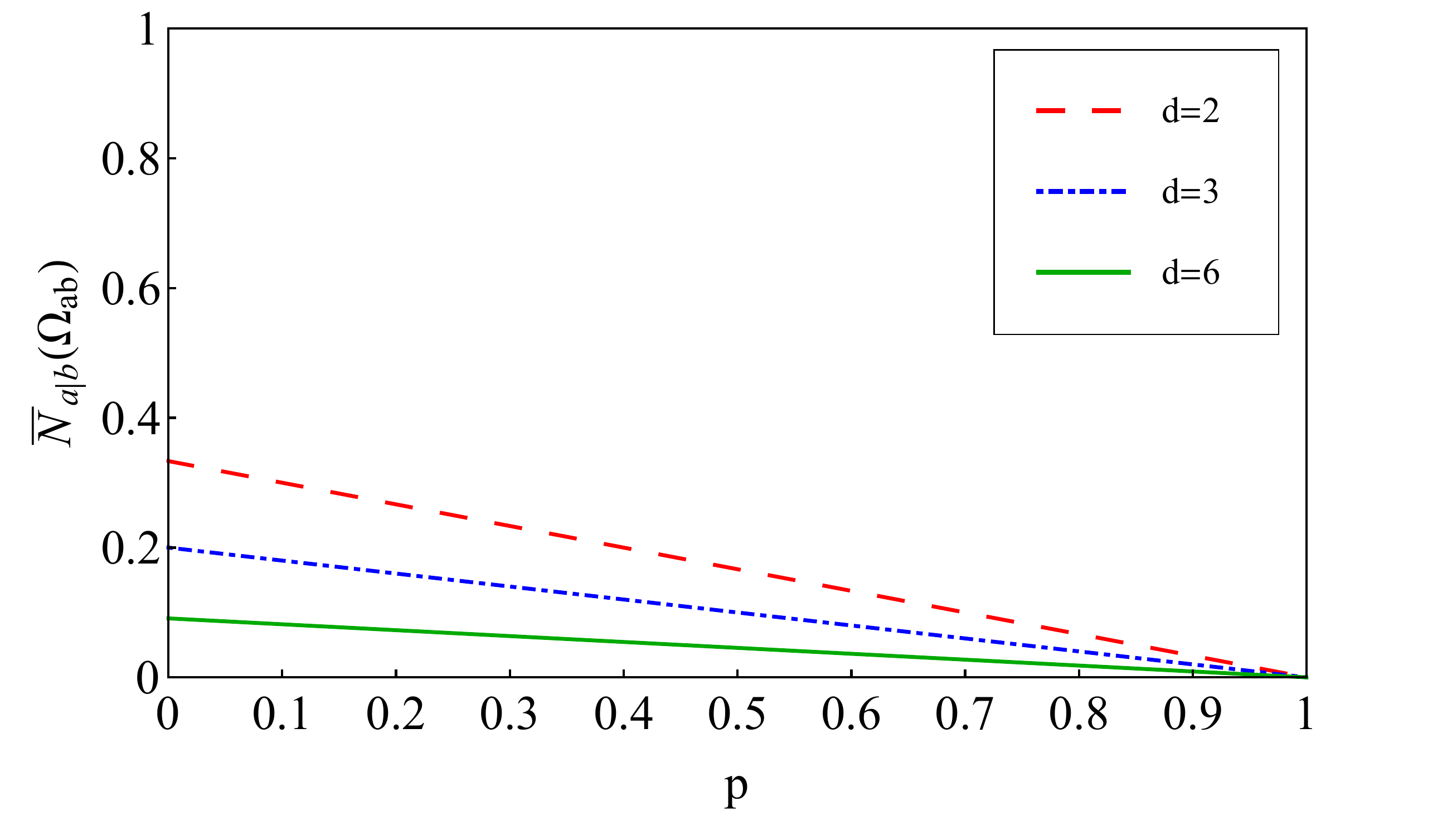} 
\caption{Average entanglement shared between qudits $a$ and $b$ (equation \eqref{DNavgAmp}), versus noise parameter $ \g $ for d = 2 (dashed
red line), d = 3 (dot dashed blue line) and d = 6 (solid green line).}
\label{fig2}
\end{figure}
is shared between qudits $a$ and $b$. Entanglement of this state is given by
\be
N_{a|b}(\Omega^{(0)}_{ab})=\frac{d(1-\g)}{d+(d-1)\g},\quad 0\leq\g\leq 1.
\ee
Hence the average shared entanglement between $a$ and $b$ is equal to:
\be\label{DNavgAmp}
\bar{N}_{a|b}(\Omega_{ab})=p_{0}N_{a|b}(\Omega^{(0)}_{ab})=\frac{1-\g}{2d-1}.
\ee
Furthermore the block-diagonal form of state in equation \eqref{D-ADOmega2'} and the same reasoning of section \ref{Noise2-qubit} results that
\be
N_{a|bc}(\Omega^{(2)'}_{abc})=N_{b|ac}(\Omega^{(2)'}_{abc})=N_{a|bc}(\Omega^{(1)'}_{abc})=\bar{N}_{a|b}(\Omega_{ab}).
\ee
Hence while exchange particle is always in separable state with rest of the system, since distillable entanglement between partitions $a|bc$ and $b|ac$ is not vanishing due to the noise, it is possible to distribute distillable entanglement between distant qudits by probabilistic EDSS protocol. Figure (\ref{fig2}) shows average distillable entanglement  shared between qudits $a$ and $b$ versus noise 
parameter $\gamma$ for $d=2$ (red dashed line), $d=3$ (blue dash-dotted line) and $d=6$ (green solid line). For amplitude damping channel we see that as the dimension of the Hilbert state increases, the amount of average entanglement distributed between qudits $a$ and $b$ decreases.  

\end{document}